\documentclass[11pt,english]{article}
\usepackage{geometry}
\usepackage{hyperref}
\makeatletter
\newcommand{\lyxaddress}[1]{
\par {\raggedright #1
\vspace{1.4em}
\noindent\par}
}

\@ifundefined{date}{}{\date{}}
\makeatother

\usepackage{babel}
\begin{document}

\title{\textbf{Surface plasmons on a thin film topological insulator}}

\author{Zohreh Davoudi%
\thanks{\href{mailto:davoudi@uw.edu}{\nolinkurl{davoudi@uw.edu}}%
}, Andreas Karch%
\thanks{\href{mailto:karch@phys.whashington.edu}{\nolinkurl{karch@phys.whashington.edu}}%
}}

\maketitle

\lyxaddress{\begin{center}
Department of Physics, University of Washington, Seattle, WA, 98195-1560,
USA
\par\end{center}}
\abstract{Recently it has been shown that surface plasmons supported by an interface between a 3+1 dimensional topological insulator and a metal or between a 3+1 dimensional topological insulator with residual bulk charge carriers and vacuum have a polarization that is rotated by an angle of order of the fine structure constant $\alpha$ compared to their topological trivial counterparts. In this work we generalize this analysis to the more realistic case of thin films, taking into account the effect of multiple reflections. In the symmetric case of a thin film surrounded by the same material on both sides the polarization of both allowed surface plasmon modes is unchanged from the single interface case, even though their dispersion relation is altered by the interactions between the two surface layers. In the general asymmetric case the angle is affected as well. We give a simple analytic expression for the resonance condition that determines the angle and solve it perturbatively for the case of a thin film between two media with almost identical permittivity.}

\section{Introduction}

Recently the properties of surface plasmons propagating at the interfaces involving three dimensional topological insulators (TIs) have been analyzed in \cite{key-3}, following a similar earlier discussion of the two dimensional case in \cite{RCQZ}.  To allow a non-trivial surface plasmon at least one of the two materials has to have a negative permittivity $\epsilon(\omega)$ as would, for example, be found in a free Drude metal. One scenario in which this can be realized is an interface between a TI and an ordinary metal. A second scenario described in \cite{key-3} is an interface between an ordinary insulator (e.g. vacuum) and a topological ``insulator" with a residual bulk charge carrier density (a situation that experimentally is much easier to achieve than a genuine TI). The properties of the surface plasmon are very similar in the two cases.

The analysis in \cite{key-3} used the TI's description in terms of their low energy effective theory. In the presence of a time reversal breaking perturbation, such as an external magnetic field, TIs are believed to be described by a topological field theory containing an $\vec{E} \cdot \vec{B}$ term
\cite{qhz}. Such a term has interesting effects on the propagation of the surface plasmons. While their dispersion relation is only modified at order $\alpha^2$ (where $\alpha$ denotes the fine structure constant), the surface plasmon's polarization is already affected at order $\alpha$. In particular, while a surface plasmon mode
on an ordinary metal/insulator interface is purely polarized in the transverse magnetic (TM) direction, it was found that for a topologically non-trivial interface the polarization is rotated by a non-vanishing angle $\nu_p$ of order $\alpha$ into the transverse electric (TE) direction. This effect is very similar to the rotation in polarization experienced by light transmitted through or reflected by a TI/vacuum interface (the Faraday and Kerr angles \cite{qhz}) and, like the latter, is potentially accessible experimentally.

The analysis in \cite{key-3} was performed only for an interface between two semi-infinite regions filled by two different materials. Of course in reality TIs will only occupy finite regions of space. In a finite geometry such as for example a thin film, the effects of multiple reflections can potentially significantly alter the properties of electromagnetic waves interacting with the interface. For example, for the case of the Faraday and Kerr reflections it was found in
\cite{Maciejko:2010yg} that the thin film geometry allows for several new effects. With this motivation in mind, we analyze the properties of a surface plasmon propagating on a thin film with two topologically non-trivially interfaces.  Our main finding is that the rotation of the plasmon's polarization (the angle $\nu_p$  mentioned above) remains completely unaffected for a thin film sandwiched in between one and the same medium on both sides, while it changes if the upper medium and the lower medium are different.

The organization of this work is as follows: in section 2 we will review the properties of surface plasmons in thin films for the case of standard insulators. In section 3 we will then generalize these findings to the case of topologically non-trivially interfaces.

\section{Review of surface plasmons on a thin film}

The basic conditions for having collective excitations of surface modes,
the quanta of which are known as surface plasmons, has been reviewed in
\cite{key-1}. In particular, we are interested in p-polarized surface
plasmons (transverse magnetic modes) which are practically accessible.
We require the plasmons to oscillate tangent to the surface and decay
exponentially perpendicular to it from both sides:
\begin{equation}
\begin{array}{cc}
\vec{H}_{\mathrm{I}}=\hat{e}_{y}H_{\mathrm{I}}^{0}e^{i(k_{\mathrm{I}}^{x}x-\omega t)}e^{ik_{\mathrm{I}}^{z}z} & z>0\\
\\
\vec{H}_{\mathrm{II}}=\hat{e}_{y}H_{\mathrm{II}}^{0}e^{i(k_{\mathrm{II}}^{x}x-\omega t)}e^{-ik_{\mathrm{II}}^{z}z} & z<0
\end{array}\label{eq: 1}
\end{equation}
and:
\begin{equation}
\begin{array}{cc}
\vec{E}_{\mathrm{I}}=\frac{H_{\mathrm{I}}^{0}}{\omega\epsilon_{\mathrm{I}}}(\hat{e}_{x}k_{\mathrm{I}}^{z}-\hat{e}_{z}k_{\mathrm{I}}^{x})e^{i(k_{\mathrm{I}}^{x}x-\omega t)}e^{ik_{\mathrm{I}}^{z}z} & z>0\\
\\
\vec{E}_{\mathrm{II}}=\frac{H_{\mathrm{II}}^{0}}{\omega\epsilon_{\mathrm{II}}}(-\hat{e}_{x}k_{\mathrm{II}}^{z}-\hat{e}_{z}k_{\mathrm{II}}^{x})e^{i(k_{\mathrm{II}}^{x}x-\omega t)}e^{-ik_{\mathrm{II}}^{z}z} & z<0\\
\\
\end{array}\label{eq: 2}
\end{equation}
Where: $k_{\mathrm{I}}^{z}\equiv i\kappa_{\mathrm{I}}$ and $k_{\mathrm{II}}^{z}\equiv i\kappa_{\mathrm{II}}$.
Note that $\kappa_{\mathrm{I}}$ and $\kappa_{\mathrm{II}}$ need
to be real and positive to guarantee an exponential decay in the normal
direction. These are solutions to Maxwell's equations which support
surface plasmons provided that the boundary conditions on the components
of electric and magnetic fields are satisfied at the interface $z=0$.
This leads to the following condition:
\begin{equation}
\frac{\kappa_{\mathrm{I}}}{\epsilon_{\mathrm{I}}}+\frac{\kappa_{\mathrm{II}}}{\epsilon_{\mathrm{II}}}=0\label{eq: 3}
\end{equation}
This implies that in order to get surface excitations of this type,
one of the media should have a negative permittivity.

The other possible solution to Maxwell's equations is the s-polarized
(transverse electric) mode. Then one can show that the necessary condition
for having surface plasmons is:
\begin{equation}
\frac{\kappa_{\mathrm{I}}}{\mu_{\mathrm{I}}}+\frac{\kappa_{\mathrm{II}}}{\mu_{\mathrm{II}}}=0\label{eq: 4}
\end{equation}
As it is hard to achieve a negative magnetic permittivity in practice,
we ignore TE modes from now on by assuming real positive magnetic
permittivity for both media equal to that of the vacuum.

Now let us consider a three medium configuration with two interfaces
at $z=0$ and $z=-d$. We call media in $z>0$, $-d<z<0$ and $z<-d$
regions medium $\mathit{\mathrm{I}}$, $\mathrm{II}$ and $\mathrm{III}$
respectively. The surface plasmon condition is changed due to
having two sets of boundary conditions at two interfaces.

One can
proceed to satisfy boundary conditions for solutions to the Maxwell's
equations as given above.
However, a simpler way is to look at the
reflectivity of an incoming light beam and find out the condition
for resonance of the system. At the $z=0$ interface the electric field
for p-polarized incoming, reflected and transmitted light are:
\begin{equation}
\begin{array}{cc}
\vec{E}_{\mathrm{I}}=\frac{E_{\mathrm{I}}c_{\mathrm{I}}}{\omega}(-\hat{e}_{x}k_{\mathrm{I}}^{z}-\hat{e}_{z}k_{\mathrm{I}}^{x})e^{i(k_{\mathrm{I}}^{x}x-\omega t)}e^{-ik_{\mathrm{I}}^{z}z} & z>0\\
\\
\vec{E}_{\mathrm{I}}^{\prime}=\frac{E_{\mathrm{I}}^{\prime}c_{\mathrm{I}}}{\omega}(\hat{e}_{x}k_{\mathrm{I}}^{z}-\hat{e}_{z}k_{\mathrm{I}}^{x})e^{i(k_{\mathrm{I}}^{x}x-\omega t)}e^{ik_{\mathrm{I}}^{z}z} & z>0\\
\\
\vec{E}_{\mathrm{II}}=\frac{E_{I\mathrm{I}}c_{\mathrm{II}}}{\omega}(-\hat{e}_{x}k_{\mathrm{II}}^{z}-\hat{e}_{z}k_{\mathrm{II}}^{x})e^{i(k_{\mathrm{II}}^{x}x-\omega t)}e^{-ik_{\mathrm{II}}^{z}z} & z<0
\end{array}\label{eq: 5}
\end{equation}
Note that for the surface plasmon we are again interested in $k_I^z=i\kappa_1$ with real and positive $\kappa$. In this case requiring a solution that has no exponentially growing piece at $z>0$ means that we are looking for a solution with only reflected, no incoming wave. Even though these names are somewhat misleading in this case as the waves aren't propagating in the $z$-direction, this is still mathematically equivalent to asking for a resonance where the total reflection coefficient, including the effect of multiple reflections, goes to infinity.

Continuity of $E_{\Vert}$ and $H_{\Vert}$ provide two equations
from which we can read the ratio of the reflected amplitude as well
as transmitted amplitude to the incoming amplitude. To proceed, we
need only the reflection coefficient from medium $\mathrm{II}$:
\begin{equation}
r_{12}\equiv\frac{E_{\mathrm{I}}^{\prime}}{E_{\mathrm{I}}}=\frac{\epsilon_{\mathrm{II}}k_{\mathrm{I}}^{z}-\epsilon_{\mathrm{I}}k_{\mathrm{II}}^{z}}{\epsilon_{\mathrm{II}}k_{\mathrm{I}}^{z}+\epsilon_{\mathrm{I}}k_{\mathrm{II}}^{z}}\label{eq: 6}
\end{equation}

To find the total reflectivity of the system, we should add up the
first reflected ray at the upper boundary and the subsequent rays
which have partially reflected from the lower and upper boundaries
and then transmitted to the medium $\mathrm{I}$ afterwards. For a
symmetric configuration, when an infinitely long thin layer of thickness
$d$ with permittivity $\epsilon_{\mathrm{II}}$ is placed in a medium
with permittivity $\epsilon_{\mathrm{I}}$, the total reflectivity $R_{12}$ of
the system can be obtained by summing the geometric series corresponding to multiple reflections:
\begin{equation}
R_{12}=\frac{r_{12}+r_{21}e^{-2\kappa_{\mathrm{II}}d}}{1-r_{21}r_{21}e^{-2\kappa_{\mathrm{II}}d}}\label{eq: 7}
\end{equation}

A resonance would occur when the denominator of total reflectivity
meets its zero; this gives the dispersion relations of the symmetric
layer system \cite{key-2}:
\begin{equation}
\begin{array}{c}
\epsilon_{\mathrm{I}}k_{\mathrm{II}}^{z}+\epsilon_{\mathrm{II}}k_{\mathrm{I}}^{z}\tanh(\frac{\kappa_{\mathrm{II}}d}{2})=0\\
\\
\epsilon_{\mathrm{I}}k_{\mathrm{II}}^{z}+\epsilon_{\mathrm{II}}k_{\mathrm{I}}^{z}\coth(\frac{\kappa_{\mathrm{II}}d}{2})=0
\end{array}\label{eq: 8}
\end{equation}
Either one of these dispersion relations describes an allowed mode. Not surprisingly, we get two surface plasmons, one for each interface. In the $d\rightarrow\infty$, we restore two copies of the surface plasmon
condition for two media (\ref{eq: 3}). For small $d$ they mix with each other, which is reflected in the modified dispersion relation.

\section{Surface plasmons on a thin film topological insulator}

In the case of a topological insulator while the Maxwell's equations
remain unchanged, the constituent relations are modified. In SI units
:
\begin{equation}
\begin{array}{c}
\vec{D}=\epsilon\vec{E}-\epsilon_{0}\alpha\frac{\theta}{\pi}(c_{0}\vec{B})\\
\\
c_{0}\vec{H}=\frac{c_{0}\vec{B}}{\mu}+\alpha\frac{\theta}{\pi}\frac{\vec{E}}{\mu_{0}}
\end{array}\label{eq: 9}
\end{equation}
where $c_{0}$ is the vacuum speed of light and $\alpha$ is the electromagnetic
structure constant. The parameter $\theta$ can take values $0$ (topologically
trivial medium) or $\pi$ (topologically non-trivial medium).

As it is shown in \cite{key-3}, the surface plasmons at the interface
of a topological insulator and a normal medium are not only TM polarized,
but they develop a non-zero TE component such that at $\mathcal{O}(\alpha)$
the ratio of the TE to TM amplitude is proportional to $\frac{\alpha\left(\theta_{\mathrm{I}}-\theta_{\mathrm{II}}\right)}{\pi}$.
So we need to write the most general form of surface plasmons' electric
and magnetic fields as a linear combination of TM and TE modes:
\begin{equation}
\begin{array}{cc}
\vec{E}_{\mathrm{I}}=-\hat{e}_{y}E_{\mathrm{I}}^{0}+\frac{E_{\mathrm{I}}c_{\mathrm{I}}}{\omega}(\hat{e}_{x}k_{\mathrm{I}}^{z}-\hat{e}_{z}k_{\mathrm{I}}^{x})e^{i(k_{\mathrm{I}}^{x}x-\omega t)}e^{ik_{\mathrm{I}}^{z}z} & z>0\\
\\
\vec{E}_{\mathrm{II}}=-\hat{e}_{y}E_{\mathrm{II}}^{0}+\frac{E_{I\mathrm{I}}c_{\mathrm{II}}}{\omega}(-\hat{e}_{x}k_{\mathrm{II}}^{z}-\hat{e}_{z}k_{\mathrm{II}}^{x})e^{i(k_{\mathrm{II}}^{x}x-\omega t)}e^{-ik_{\mathrm{II}}^{z}z} & z<0
\end{array}\label{eq: 10}
\end{equation}
Using the relation: $\frac{\vec{E}}{c}=-\frac{\vec{k}}{\omega}\times(c\vec{B})$
gives:
\begin{equation}
\begin{array}{cc}
\vec{B}_{\mathrm{I}}=\hat{e}_{y}\frac{E_{\mathrm{I}}}{c_{\mathrm{I}}}+\frac{E_{\mathrm{I}}^{0}}{\omega}(\hat{e}_{x}k_{\mathrm{I}}^{z}-\hat{e}_{z}k_{\mathrm{I}}^{x})e^{i(k_{\mathrm{I}}^{x}x-\omega t)}e^{ik_{\mathrm{I}}^{z}z} & z>0\\
\\
\vec{B}_{\mathrm{II}}=\hat{e}_{y}\frac{E_{\mathrm{II}}}{c_{\mathrm{II}}}+\frac{E_{\mathrm{II}}^{0}}{\omega}(-\hat{e}_{x}k_{\mathrm{II}}^{z}-\hat{e}_{z}k_{\mathrm{II}}^{x})e^{i(k_{\mathrm{II}}^{x}x-\omega t)}e^{-ik_{\mathrm{II}}^{z}z} & z<0
\end{array}\label{eq: 11}
\end{equation}

Using the constituent relations (\ref{eq: 9}) in both media, we
can easily build $\vec{H}$ and $\vec{D}$ fields. Then boundary conditions
on the field components at $z=0$ give the dispersion relation as well
as TE to TM amplitude ratio. The former as is shown in \cite{key-3}
is only modified at $\mathcal{O}(\alpha^{2})$ while the later gets
modification at $\mathcal{O}(\alpha)$:
\begin{equation}
\tan(\nu_{p})\equiv\frac{E_{\mathrm{II}}^{0}}{E_{II}}=-\frac{\alpha\triangle\theta}{\pi}\frac{c_{\mathrm{II}}\epsilon_{\mathrm{II}}}{c_{0}(\epsilon_{\mathrm{I}}-\epsilon_{\mathrm{II}})}+\mathcal{O}((\frac{\alpha\triangle\theta}{\pi})^{2})\label{eq: 12}
\end{equation}
where: $\triangle\theta=\theta_{\mathrm{I}}-\theta_{\mathrm{II}}$
and we should keep in mind that $\epsilon_{\mathrm{I}}\epsilon_{\mathrm{II}}<0$
and: $\epsilon_{\mathrm{I}}+\epsilon_{\mathrm{II}}<0$ \cite{key-3}.

We aim to see how the plasmon polarization angle $\nu_{p}$ will change
in case of a thin layer of topological insulator placed in a normal
medium.
Similar to the non-topological thin film, we analyze the problem by
looking at the resonance state of a light beam reflecting from the
film. The incoming, reflected and transmitted electric and magnetic
fields at the boundary $z=0$ are respectively:
\begin{equation}
\begin{array}{cc}
\vec{E}_{\mathrm{I}}=-\hat{e}_{y}E_{\mathrm{I}}^{0}+\frac{E_{\mathrm{I}}c_{\mathrm{I}}}{\omega}(-\hat{e}_{x}k_{\mathrm{I}}^{z}-\hat{e}_{z}k_{\mathrm{I}}^{x})e^{i(k_{\mathrm{I}}^{x}x-\omega t)}e^{-ik_{\mathrm{I}}^{z}z} & z>0\\
\\
\vec{E}_{\mathrm{I}}^{\prime}=-\hat{e}_{y}E_{\mathrm{I}}^{0\prime}+\frac{E_{\mathrm{I}}^{\prime}c_{\mathrm{I}}}{\omega}(\hat{e}_{x}k_{\mathrm{I}}^{z}-\hat{e}_{z}k_{\mathrm{I}}^{x})e^{i(k_{\mathrm{I}}^{x}x-\omega t)}e^{ik_{\mathrm{I}}^{z}z} & z>0\\
\\
\vec{E}_{\mathrm{II}}=-\hat{e}_{y}E_{\mathrm{II}}^{0}+\frac{E_{I\mathrm{I}}c_{\mathrm{II}}}{\omega}(-\hat{e}_{x}k_{\mathrm{II}}^{z}-\hat{e}_{z}k_{\mathrm{II}}^{x})e^{i(k_{\mathrm{II}}^{x}x-\omega t)}e^{-ik_{\mathrm{II}}^{z}z} & z<0
\end{array}\label{eq: 13}
\end{equation}
and:
\begin{equation}
\begin{array}{cc}
\vec{B}_{\mathrm{I}}^{\prime}=\hat{e}_{y}\frac{E_{\mathrm{I}}}{c_{\mathrm{I}}}+\frac{E_{\mathrm{I}}^{0}}{\omega}(-\hat{e}_{x}k_{\mathrm{I}}^{z}-\hat{e}_{z}k_{\mathrm{I}}^{x})e^{i(k_{\mathrm{I}}^{x}x-\omega t)}e^{-ik_{\mathrm{I}}^{z}z} & z>0\\
\\
\vec{B}_{\mathrm{I}}=\hat{e}_{y}\frac{E_{\mathrm{I}}^{\prime}}{c_{\mathrm{I}}}+\frac{E_{\mathrm{I}}^{0\prime}}{\omega}(\hat{e}_{x}k_{\mathrm{I}}^{z}-\hat{e}_{z}k_{\mathrm{I}}^{x})e^{i(k_{\mathrm{I}}^{x}x-\omega t)}e^{ik_{\mathrm{I}}^{z}z} & z>0\\
\\
\vec{B}_{\mathrm{II}}=\hat{e}_{y}\frac{E_{\mathrm{II}}}{c_{\mathrm{II}}}+\frac{E_{\mathrm{II}}^{0}}{\omega}(-\hat{e}_{x}k_{\mathrm{II}}^{z}-\hat{e}_{z}k_{\mathrm{II}}^{x})e^{i(k_{\mathrm{II}}^{x}x-\omega t)}e^{-ik_{\mathrm{II}}^{z}z} & z<0
\end{array}\label{eq: 14}
\end{equation}

Using the constituent relations (\ref{eq: 9}), we can also form $\vec{H}$
and $\vec{D}$ fields. Boundary conditions are the continuity of $\vec{E}_{\Vert}$
and $\vec{H}_{\Vert}$ as well as continuity of $\vec{D}_{\bot}$
and $\vec{B}_{\bot}$ at $z=0$. This gives six equations, out of
which two are redundant. For any of the continuity equations to hold for all $x$ and $t$ we need
 $k_{\mathrm{I}}^{x}$ to be equal to $k_{\mathrm{II}}^{x}$. So we
are left with four independent linear equations with six unknowns:
$E_{\mathrm{I}}^{0}$, $E_{\mathrm{I}}$, $E_{\mathrm{I}}^{0\prime}$,
$E_{\mathrm{I}}^{\prime}$, $E_{\mathrm{II}}^{0}$ and $E_{\mathrm{II}}$.

Treating $E_{\mathrm{I}}^{0}$ and $E_{\mathrm{I}}$, that is the amplitudes of the two polarization components of the incoming wave, as input parameters, we can solve for the corresponding amplitudes of the reflected light in terms of a reflection matrix $R$ defined as:
\begin{equation}
\left(\begin{array}{c}
E_{\mathrm{I}}^{0\prime}\\
E_{\mathrm{I}}^{\prime}
\end{array}\right)=R_{12}\left(\begin{array}{c}
E_{\mathrm{I}}^{0}\\
E_{\mathrm{I}}
\end{array}\right).
\end{equation}
That is the reflected amplitudes $E_{\mathrm{I}}^{0\prime}$ and $E_{\mathrm{I}}^{\prime}$
are linear combinations of the incoming ones, $E_{\mathrm{I}}^{0}$ and $E_{\mathrm{I}}$;
if one sends a purely TM (or TE) polarized light toward a topological
insulator, the reflected light has mixed TM and TE components. Explicitly we find
\begin{equation}
\begin{array}{c}
E_{\mathrm{I}}^{0\prime}=\frac{-E_{\mathrm{I}}^{0}\left[-\epsilon_{\mathrm{II}}(k_{\mathrm{I}}^{z})^{2}+\epsilon_{\mathrm{I}}(k_{\mathrm{II}}^{z})^{2}+k_{\mathrm{I}}^{z}k_{\mathrm{II}}^{z}\left(-\epsilon_{\mathrm{I}}+\epsilon_{\mathrm{II}}+\epsilon_{0}(\frac{\alpha\triangle\theta}{\pi})^{2}\right)\right]+E_{\mathrm{I}}\left[\frac{2\epsilon_{\mathrm{I}}c_{\mathrm{I}}}{c_{0}}(\frac{\alpha\triangle\theta}{\pi})k_{\mathrm{I}}^{z}k_{\mathrm{II}}^{z}\right]}{\epsilon_{\mathrm{II}}(k_{\mathrm{I}}^{z})^{2}+\epsilon_{\mathrm{I}}(k_{\mathrm{II}}^{z})^{2}+k_{\mathrm{I}}^{z}k_{\mathrm{II}}^{z}\left(\epsilon_{\mathrm{I}}+\epsilon_{\mathrm{II}}+\epsilon_{0}(\frac{\alpha\triangle\theta}{\pi})^{2}\right)}\\
\\
E_{\mathrm{I}}^{\prime}=\frac{E_{\mathrm{I}}^{0}\left[\frac{2\epsilon_{\mathrm{I}}c_{\mathrm{I}}}{c_{0}}(\frac{\alpha\triangle\theta}{\pi})k_{\mathrm{I}}^{z}k_{\mathrm{II}}^{z}\right]+E_{\mathrm{I}}\left[\epsilon_{\mathrm{II}}(k_{\mathrm{I}}^{z})^{2}-\epsilon_{\mathrm{I}}(k_{\mathrm{II}}^{z})^{2}+k_{\mathrm{I}}^{z}k_{\mathrm{II}}^{z}\left(-\epsilon_{\mathrm{I}}+\epsilon_{\mathrm{II}}+\epsilon_{0}(\frac{\alpha\triangle\theta}{\pi})^{2}\right)\right]}{\epsilon_{\mathrm{II}}(k_{\mathrm{I}}^{z})^{2}+\epsilon_{\mathrm{I}}(k_{\mathrm{II}}^{z})^{2}+k_{\mathrm{I}}^{z}k_{\mathrm{II}}^{z}\left(\epsilon_{\mathrm{I}}+\epsilon_{\mathrm{II}}+\epsilon_{0}(\frac{\alpha\triangle\theta}{\pi})^{2}\right)}
\end{array}\label{eq: 17}
\end{equation}
where we have set $\mu_{\mathrm{I}}=\mu_{\mathrm{II}}=\mu_{0}$.

For the symmetric case of a thin TI film immersed in a medium (so that $\epsilon_{\mathrm{I}}=\epsilon_{\mathrm{III}}$ and similar for $\theta$ and $\mu$), it is most convenient to first diagonalize the matrix $R$. An eigenvector of $R$ corresponds to a polarization for which the incoming light gets reflected off the interface without being rotated. In terms of these eigenvectors we can then perform the resonance analysis as in the topological trivial thin film. We'll find two different resonance conditions for the two different eigenvectors. One of them is the rotated TM mode we are interested in. That is, its polarization only differs from TM by order $\alpha$ and it needs negative $\epsilon$ in one of the media to be supported by the interface. The second polarization is the rotated TE mode. Its resonance condition can not be met unless one
of the $\mu$'s is negative and so we'll disregard it as before.
Naming the eigenvalues of the matrix $R_{12}$ as $\left(\tilde{r_{0}}\right)_{12}$
and $\left(\tilde{r}\right)_{12}$, the well-defined reflection coefficients
then are:
\begin{equation}
\begin{array}{c}
\left(\tilde{r_{0}}\right)_{12}=\frac{\epsilon_{\mathrm{II}}(k_{\mathrm{I}}^{z})^{2}-\epsilon_{\mathrm{I}}(k_{\mathrm{II}}^{z})^{2}-k_{\mathrm{I}}^{z}k_{\mathrm{II}}^{z}\sqrt{\epsilon_{0}^{2}(\frac{\alpha\triangle\theta}{\pi})^{4}+2\epsilon_{0}(\epsilon_{\mathrm{I}}+\epsilon_{\mathrm{II}})(\frac{\alpha\triangle\theta}{\pi})^{2}+(\epsilon_{\mathrm{I}}-\epsilon_{\mathrm{II}})^{2}}}{\epsilon_{\mathrm{II}}(k_{\mathrm{I}}^{z})^{2}+\epsilon_{\mathrm{I}}(k_{\mathrm{II}}^{z})^{2}+k_{\mathrm{I}}^{z}k_{\mathrm{II}}^{z}\left(\epsilon_{\mathrm{I}}+\epsilon_{\mathrm{II}}+\epsilon_{0}(\frac{\alpha\triangle\theta}{\pi})^{2}\right)}\\
\\
\left(\tilde{r}\right)_{12}=\frac{\epsilon_{\mathrm{II}}(k_{\mathrm{I}}^{z})^{2}-\epsilon_{\mathrm{I}}(k_{\mathrm{II}}^{z})^{2}+k_{\mathrm{I}}^{z}k_{\mathrm{II}}^{z}\sqrt{\epsilon_{0}^{2}(\frac{\alpha\triangle\theta}{\pi})^{4}+2\epsilon_{0}(\epsilon_{\mathrm{I}}+\epsilon_{\mathrm{II}})(\frac{\alpha\triangle\theta}{\pi})^{2}+(\epsilon_{\mathrm{I}}-\epsilon_{\mathrm{II}})^{2}}}{\epsilon_{\mathrm{II}}(k_{\mathrm{I}}^{z})^{2}+\epsilon_{\mathrm{I}}(k_{\mathrm{II}}^{z})^{2}+k_{\mathrm{I}}^{z}k_{\mathrm{II}}^{z}\left(\epsilon_{\mathrm{I}}+\epsilon_{\mathrm{II}}+\epsilon_{0}(\frac{\alpha\triangle\theta}{\pi})^{2}\right)}
\end{array}\label{eq: 18}
\end{equation}

In order to impose two resonance conditions:
\begin{equation}
\begin{array}{c}
1-(\tilde{r_{0}})_{21}(\tilde{r_{0}})_{21}e^{-2\kappa_{\mathrm{II}}d}=0\\
\\
1-(\tilde{r})_{21}(\tilde{r})_{21}e^{-2\kappa_{\mathrm{II}}d}=0
\end{array}\label{eq: 19}
\end{equation}
we need $\left(\tilde{r_{0}}\right)_{21}$ and $\left(\tilde{r}\right)_{21}$
coefficients instead. These are eigenvalues of matrix $R_{21}$ obtained from (\ref{eq: 17})
after interchanging $\mathrm{I}$ and $\mathrm{II}$ indices.
These conditions give two dispersion relations. Defining the variable
$x\equiv\frac{k_{\mathrm{I}}^{z}}{k_{\mathrm{II}}^{z}}$, it is not
hard to see that these relations are a polynomial equation quartic
in $x$ and has been modified by terms of at least $\mathcal{O}(\alpha^{2})$
similar to the result obtained in \cite{key-3} for the topologically
non-trivial two medium system. Now let us assume that medium $\mathrm{I}$
is a normal insulator such as vacuum, and medium $\mathrm{II}$ is
a topological insulator, then $\epsilon_{\mathrm{II}}$ is necessarily
negative and also: $\left|\epsilon_{\mathrm{II}}\right|>\epsilon_{\mathrm{I}}$
as stated in Ref {[}3{]}. Then one can show that the only solutions
for the variable $x$ which reduce to (\ref{eq: 8}) at $\triangle\theta=0$
come from the second equation in (\ref{eq: 19}). These are the rotated TM modes we are looking for; the other resonance condition would correspond to the rotated TE modes.

In order to extract the angle $\nu_{p}$ we need to look at the ratio of the TE to the TM amplitude
for the rotated TM mode. To do so, we should look
at the corresponding eigenvector of the matrix $R_{21}$ as defined above and apply
the dispersion relation to its components. But one can easily check
that these eigenvectors are independent of $x$, and we have:
\begin{equation}
\tan(\nu_{p})\equiv\frac{E_{\mathrm{II}}^{0}}{E_{\mathrm{II}}}=\frac{\epsilon_{\mathrm{I}}-\epsilon_{\mathrm{II}}+\epsilon_{0}(\frac{\alpha\triangle\theta}{\pi})^{2}-\sqrt{\left(\epsilon_{\mathrm{I}}-\epsilon_{\mathrm{II}}+\epsilon_{0}(\frac{\alpha\triangle\theta}{\pi})^{2}\right)^{2}+\left(\frac{2\epsilon_{\mathrm{II}}c_{\mathrm{II}}}{c_{0}}\left(\frac{\alpha\triangle\theta}{\pi}\right)\right)^{2}}}{\frac{2\epsilon_{\mathrm{II}}c_{\mathrm{II}}}{c_{0}}(\frac{\alpha\triangle\theta}{\pi})}\label{eq: 20}
\end{equation}

Expanding equation (\ref{eq: 20}) up to $\mathcal{O}(\alpha^{2})$
gives:
\begin{equation}
\begin{array}{ccc}
\tan(\nu_{p})=-\frac{\alpha\triangle\theta}{\pi}\frac{c_{\mathrm{II}}\epsilon_{\mathrm{II}}}{c_{0}(\epsilon_{\mathrm{I}}-\epsilon_{\mathrm{II}})}+\mathcal{O}((\frac{\alpha\triangle\theta}{\pi})^{2}) &  & \epsilon_{\mathrm{I}}>\epsilon_{\mathrm{II}}\end{array}\label{eq: 21}
\end{equation}
which shows that $\tan(\nu_{p})$ will not change in case of a thin
film topological insulator placed in a normal medium compared
to a semi-infinite topological insulator in proximity to that normal
medium.

This result seems to be a consequence of the symmetric configuration we
have considered. If the upper and lower layers are filled with
different substances, we should expect a different polarization angle.
To prove this claim, we proceed as follows. The general resonance
condition should be written in the matrix form:
\begin{equation}
1_{2\times2}-R_{21}R_{23}e^{-2\kappa d}=0\label{eq: 22}
\end{equation}
Note that $R_{21}$ and $R_{23}$ can, in general, not be diagonalized simultaneously,
so we need to diagonalize the matrix $M\equiv1_{2\times2}-R_{21}R_{23}e^{-2\kappa d}$
as a whole and finding the dispersion relations by setting its eigenvalues
equal to zero. Although this can be done numerically quite easily,
a concise analytic solution for the general asymmetric case can not
be obtained. But in order to show how the polarization angle is affected
in this configuration analytically, we adopt a simple asymmetric case
where the top layer is vacuum and the bottom layer is only slightly different
from vacuum while the middle layer is a topological insulator:
\begin{equation}
\begin{array}{c}
\epsilon_{\mathrm{I}}=\epsilon_{0}=1\\
\epsilon_{\mathrm{II}}=\epsilon\\
\epsilon_{\mathrm{III}}=\epsilon_{0}\left(1+\delta\right)=1+\delta
\end{array}\label{eq: 23}
\end{equation}
With $\delta\ll1$. And we set $c_{0}=1$ and $c_{\mathrm{II}}=c$.
Also:
\begin{equation}
\begin{array}{c}
x\equiv\frac{k_{\mathrm{I}}^{z}}{k_{\mathrm{II}}^{z}}=x_{0}+\delta\, x_{1}\\
\\
y\equiv\frac{k_{\mathrm{III}}^{z}}{k_{\mathrm{II}}^{z}}=y_{0}+\delta\, y_{1}
\end{array}\label{eq: 24}
\end{equation}
where $x_{0}=y_{0}$ is the solution to the symmetric resonance equation.
We can plug these values into the eigenvalues and eigenvectors of
matrix $M$ and expand them at order $\delta$. Although the eigenvalues, which determine the dispersion relation after being set equal to zero,
depend on $x_{1}$ and $y_{1}$ explicitly, the eigenvectors on the other hand are independent
of $x_{1}$ and $y_{1}$ at this order and we only need to use the
solution to $x_{0}=y_{0}$ of the symmetric case, which we found before. Note that there are two acceptable solutions for $x_{0}$ which
reduce to the two modes of (\ref{eq: 8}) at $\Delta\theta=0$. That is they correspond to the expected two rotated TM modes supported by the two interfaces. One can show
that there are two corresponding polarization angle at $\mathcal{O}\left(\alpha\delta\right)$
as following:
\begin{equation}
\begin{array}{c}
\tan(\nu_{p})=-\frac{\alpha\triangle\theta}{\pi}\frac{c\epsilon}{(1-\epsilon)}\left[1+\frac{1}{(1-\epsilon)}\frac{\left(1+e^{\kappa d}\right)+\epsilon\left(-1+e^{\kappa d}\right)}{-\left(1+e^{\kappa d}\right)^{2}+\epsilon\left(-1+e^{\kappa d}\right)^{2}}\delta+\mathcal{O}\left(\delta^{2}\right)\right]+\mathcal{O}\left(\alpha^{2}\right)\\
\\
\tan(\nu_{p})=-\frac{\alpha\triangle\theta}{\pi}\frac{c\epsilon}{(1-\epsilon)}\left[1-\frac{1}{(1-\epsilon)}\frac{\left(-1+e^{\kappa d}\right)+\epsilon\left(1+e^{\kappa d}\right)}{-\left(-1+e^{\kappa d}\right)^{2}+\epsilon\left(1+e^{\kappa d}\right)^{2}}\delta+\mathcal{O}\left(\delta^{2}\right)\right]+\mathcal{O}\left(\alpha^{2}\right)
\end{array}\label{eq: 25}
\end{equation}
where $\epsilon<1$ and $\left|\epsilon_{\mathrm{II}}\right|>\epsilon_{\mathrm{I}}$.

This clearly shows that if we place a thin film topological insulator
between two different media, the polarization angle $\nu_{p}$ changes
compared with a configuration where a thin film topological insulator
is placed in one medium.

\section*{Acknowledgements}
This work was supported in part by the U.S. Department of Energy
under Grant No. DE-FG02-96ER40956.

\end{document}